\documentclass[conference]{IEEEtran}
\IEEEoverridecommandlockouts
\usepackage{cite}
\usepackage{amsmath,amssymb,amsfonts}
\usepackage{algorithmic}
\usepackage{graphicx}
\usepackage{textcomp}
\usepackage{xcolor}
\usepackage{subcaption}
\usepackage{amsmath}
\usepackage{tikz}
\usetikzlibrary{quantikz}
\usepackage[ruled,vlined]{algorithm2e}
\usepackage{booktabs}
\usepackage{comment}
\usepackage{xurl}
\usepackage{float}

\def\BibTeX{{\rm B\kern-.05em{\sc i\kern-.025em b}\kern-.08em
    T\kern-.1667em\lower.7ex\hbox{E}\kern-.125emX}}

\makeatletter

\def\equalsfill{$\m@th\mathord=\mkern-7mu
\cleaders\hbox{$\!\mathord=\!$}\hfill
\mkern-7mu\mathord=$}

\newcommand{\linebreakand}{%
  \end{@IEEEauthorhalign}
  \hfill\mbox{}\par
  \mbox{}\hfill\begin{@IEEEauthorhalign}}

\makeatother

\begin{document}

\title{Distinguishing Quantum Software Bugs from Hardware Noise: A Statistical Approach}

\author{\IEEEauthorblockN{ Ahmik Virani}
\IEEEauthorblockA{\textit{Dept of Engineering Science} \\
\textit{IIT Hyderabad}\\
Kandi, India \\
es22btech11001@iith.ac.in}
\and
\IEEEauthorblockN{Devraj}
\IEEEauthorblockA{\textit{Dept of Engineering Science} \\
\textit{IIT Hyderabad}\\
Kandi, India \\
es22btech11011@iith.ac.in}
\and
\IEEEauthorblockN{Anirudh Suresh}
\IEEEauthorblockA{\textit{Dept of Computer Science} \\
\textit{University of Maryland College Park}\\
Maryland, United States \\
asuresh3@terpmail.umd.edu}
\linebreakand
\IEEEauthorblockN{Lei Zhang}
\IEEEauthorblockA{\textit{Dept of Information Systems } \\
\textit{University of Maryland Baltimore County}\\
Maryland, United States \\
leizhang@umbc.edu}
\and
\IEEEauthorblockN{ M V Panduranga Rao}
\IEEEauthorblockA{\textit{Dept of Computer Science and Engineering} \\
\textit{IIT Hyderabad}\\
Kandi, India\\
mvp@cse.iith.ac.in}

}

\maketitle

\begin{abstract}
Quantum computing in the Noisy Intermediate-Scale Quantum (NISQ) era presents significant challenges in differentiating quantum software bugs from hardware noise. Traditional debugging techniques from classical software engineering cannot directly resolve this issue due to the inherently stochastic nature of quantum computation mixed with noises from NISQ computers. To address this gap, we propose a statistical approach leveraging probabilistic metrics to differentiate between quantum software bugs and hardware noise. We evaluate our methodology empirically using well-known quantum algorithms, including Grover’s algorithm, Deutsch-Jozsa algorithm, and Simon’s algorithm. Experimental results demonstrate the efficacy and practical applicability of our approach, providing quantum software developers with a reliable analytical tool to identify and classify unexpected behavior in quantum programs.
\end{abstract}

\section{Introduction}\label{sec:intro}

Quantum computing promises significant advancements in fields, such as cryptography, optimization, and simulation, leveraging quantum mechanics (like superposition and entanglement). Consequently, quantum development platforms (e.g., Qiskit~\cite{Qiskit-Documentation}, Q\#~\cite{QMicrosoft}, and PennyLane~\cite{PennyLane}) and quantum software (e.g., quantum simulators and quantum convolutional neural networks) have evolved dramatically in the last decade. However, quantum software engineering in the current Noisy Intermediate-Scale Quantum (NISQ) era faces critical challenges, primarily due to the presence of \textbf{quantum noise}, which significantly impacts the quality and reliability of quantum software~\cite{bharti2022noisy}. 

Unlike classical systems, where software bugs can be distinguished clearly from hardware failures, quantum software debugging faces unique difficulties in differentiating between quantum software bugs and hardware noise~\cite{pan2023understanding,Salonik}. Misclassification between these two issues can lead to considerable resource wastage. The literature lacks methods for debugging quantum software with noises due to the indeterminacy of quantum software. 
While there are techniques like quantum error correction for mitigating quantum noise~\cite{muqeet2024mitigatingnoisequantumsoftware}, its efficacy depends on what implementation technology is being used.

Specifically, it is crucial to know, as a quantum programmer, if an unexpected output obtained is because of quantum noise inherent to the hardware or because of a bug in the
quantum code itself. In this work, we refer to deciding between the four possibilities as the \emph{output dilemma}: 
\begin{enumerate}
    \item No Bugs, No Noise: the quantum program being correct and the quantum computer
being noise-free,
    \item No Bugs, Noisy: the quantum program being correct, and the quantum computer being noisy,
    \item Buggy, No Noise: the quantum program being buggy, and the quantum computer being noise-free, and
    \item Buggy, Noisy: the quantum program being incorrect, and the quantum computer
being noisy.
\end{enumerate}

To address this challenge, we propose a probabilistic approach employing statistical metrics to effectively differentiate between bugs in quantum software and hardware noise. This approach provides a systematic way to diagnose unexpected behavior in quantum programs by leveraging statistical insights derived from quantum measurements. 

A meaningful quantum algorithm will need to result in an
elevated probability of collapsing to specific eigenstates of an observable for computing the solution to a computational problem. This is essential for efficiently separating a wrong solution from a correct one. In this work, we address those quantum algorithms for which
the cardinality of such elevated probability eigenstates is known in advance through
theoretical analysis. Indeed, this property is displayed by all folklore quantum
algorithms, as will be seen in the coming sections.

The \textbf{contributions} of this paper are as follows.
\begin{enumerate}
    \item We propose a statistical method, which we call the Bias-Entropy Model, for distinguishing quantum bugs from noise.
    \item We empirically validate these metrics through representative quantum algorithms (including Grover's algorithm~\cite{grover1996fast}, Deutsch-Jozsa algorithm~\cite{deutsch1992rapid}, and Simon's algorithm~\cite{simon1994power}).
\end{enumerate}

The experimental results show the effectiveness and applicability of our proposed method. The code for our experiments is made available at https://github.com/Ahmik-Virani/Differentiating-Quantum-Bug-From-Noise-Statistical-Approach.

We use Grover's algorithm as a running example to present our ideas and approach.
Given an oracle that ``marks" some element(s) in an unordered list, Grover's quantum algorithm returns the index (indices) of the element(s) and provides a quadratic speed-up as compared to classical computing~\cite{grover1996fast}. 

Before applying the bias-entropy technique to some folklore algorithms as case-studies, we analyze through simulations
the effect of noise and bugs individually on bias and entropy. To carry out this study, we introduce noise and bugs (generated through Muskit~\cite{Mendiluze2021}) on randomly generated quantum circuits.
Such a study goes to support our intuition as to why the bias-entropy technique would be useful in analyzing quantum programs.

The remainder of the paper is structured as follows. Section~\ref{sec:prel} presents the background and related work. Section~\ref{sec:Approach} introduces our approach.  Section~\ref{sec:effects_noise_bug} discusses an empirical demonstration of the effects of bias and entropy on randomly generated quantum circuits. Section~\ref{sec:case-study} describes experimental results on circuits for folklore quantum
algorithms. Section~\ref{sec:threats} discusses threats to validity of the results obtained in this work. Section~\ref{sec:conclusions} concludes the paper with a brief discussion of future work.

\section{Related Work and Preliminaries}\label{sec:prel}

In this work, we assume familiarity with basic ideas in quantum computing---like quantum state vectors, unitary gates, particularly the Pauli gates, measurements, etc~\cite{Nielsen_Chuang_2010}. 

\subsection{Quantum Bugs and Noise}

We begin with some terminology in the context of quantum programs that we will use in this paper:

\begin{itemize} 
\item Bugs \cite{Paltenghi_2022}: \emph{A bug refers to discrepancies in the implementation of a quantum algorithm that arise from logical errors in the software code.} These errors manifest as deviations from the algorithm’s intended behavior caused by, for example, incorrect gate sequences or misconfigured parameters. Bugs stem from human error during the design or programming phase, rather than physical hardware limitations. 

\item Noise \cite{rouzé2023efficientlearningstructureparameters, cordier2025scalingproblemsizesenvironmental}: \emph{Noise refers to stochastic errors introduced during quantum computation due to hardware imperfections or simulated environmental interactions.} In physical quantum devices, noise arises from factors such as qubit decoherence, gate infidelity, and crosstalk. In quantum simulators (e.g., Qiskit Aer Simulator\cite{qiskit-aer}) noise is artificially modeled through channels like the depolarizing noise channel, which applies randomly chosen Pauli operators to qubits with some probability. 

\end{itemize}

\subsection{Related Work}

Recent research in Quantum Software Engineering has focused on defining quantum-specific software engineering methods, design patterns, and quality assurance techniques~\cite{zhao2020quantum,10.1145/3712002}. However, a gap remains in systematically identifying, classifying, and mitigating quantum software bugs, particularly those influenced by quantum hardware noise.

Huang and Martonosi~\cite{huang2019statistical} categorized quantum bugs into algorithmic, coding, and compilation issues, using statistical assertions. However, recent studies highlight the critical role of quantum hardware noise in complicating bug detection and classification, suggesting the need for noise-aware debugging techniques~\cite{Paltenghi_2022,moguel2022quantum}. 
Muqeet et al.~\cite{muqeet2024mitigatingnoisequantumsoftware} propose a noise-aware 
method using machine learning techniques to learn the effect of noise on a quantum computer and filter it from a program’s output.

Notable work has been done in the field of studying quantum bug detection and debugging. Tools such as QuanFuzz\cite{wang2018quanfuzzfuzztestingquantum}, QMutPy\cite{10.1145/3533767.3543296}, and Muskit\cite{Mendiluze2021} have been developed to analyze the impact of bugs in quantum programs and enhance debugging methodologies for quantum systems. Furthermore, fuzz testing \cite{9787950, wang2018quanfuzzfuzztestingquantum, 9438606} and mutation testing \cite{9844849, Mendiluze2021} techniques for quantum software have also been
investigated in the recent past.

Quantum noise poses a challenge for achieving the desired accuracy for a quantum program. Unlike classical computing, quantum programs are susceptible to different noise effects, causing deviation of the output of the quantum program \cite{Preskill2018quantumcomputingin}, making it non-trivial to measure the accuracy and check for correctness of the quantum program.

Compared to the previous machine learning techniques in QOIN~\cite{muqeet2024mitigatingnoisequantumsoftware}, which rely on training, our work focuses on statistical approaches, which provide an explicit understanding of quantum noise characteristics and interactions of bugs and noise. Moreover, our proposed approaches present a set of systematic techniques for identifying, classifying, and distinguishing quantum software bugs from hardware noise.

As per Ramalho et al., \cite{ramalho2024testingdebuggingquantumprograms}, current methodologies in quantum software testing often overlook the practical constraints of real quantum hardware, particularly the impact of noise on computational reliability. A critical challenge lies in distinguishing inherently faulty program behavior—stemming from algorithmic or implementation errors—from the stochastic outcomes induced by noise. To bridge this gap, we propose a method that employs a quantitative metric to establish acceptable noise thresholds. This approach allows us to assess whether the current runtime environment is suitable for testing and determine whether faults in the output stem from bugs or noise.

\subsection{Custom Noise Model Construction}\label{sec:Noise_model}

We now discuss noise models, that we also use in this work.
A custom quantum noise model can be constructed using separate depolarizing error channels\cite{math12091385, Leditzky_2018, Nielsen_Chuang_2010} for single-qubit and two-qubit gates. These depolarizing error channels introduces a Pauli error in the output of any gate operation in the circuit. For a target total error probability $p$, density matrix of the original quantum state $\rho$, and the depolarizing parameter $\lambda$, the noise model is defined as follows.

\subsubsection{Single-Qubit Gates}
For each single-qubit gate, the depolarizing channel is implemented as follows.
\begin{itemize}
\item The depolarizing channel $\mathcal{D}_{1,\lambda}$ is defined as
\begin{equation*}
\mathcal{D}_{1,\lambda}(\rho) = (1-\lambda)\rho + \frac{\lambda}{4}(X\rho X + Y\rho Y + Z\rho Z + I\rho I)
\end{equation*}
\begin{equation*}
\implies \mathcal{D}_{1,\lambda}(\rho) = (1- \frac{3 \lambda}{4})\rho + \frac{\lambda}{4}(X\rho X + Y\rho Y + Z\rho Z)
\end{equation*}
Choosing $\lambda = \cfrac{4p}{3}$ gives us a combined probability of error equal to $p$ with each Pauli error ($X$, $Y$, $Z$) occurring with probability $\cfrac{\lambda}{4} = \cfrac{p}{3}$.
\item $\lambda_{\text{full}} = \cfrac{4}{3}$ corresponds to the maximum allowed depolarizing parameter. The corresponding value of $p=1$ gives us an error channel where every single-qubit gate operation is necessarily followed by a uniformly random single-qubit Pauli error.
\end{itemize}

\subsubsection{Two-Qubit Gates}
For two-qubit gates, a depolarizing error channel is implemented as follows.

\begin{itemize}
\item The two-qubit depolarizing channel $\mathcal{D}_{2,\lambda}$ is
\begin{equation*}
\mathcal{D}_{2,\lambda}(\rho) = (1-\cfrac{15\lambda}{16})\rho + \cfrac{\lambda}{16}\sum_{P \in \mathcal{P}_2} P\rho P^\dagger,
\end{equation*}
where $\mathcal{P}_2$ contains all 15 non-identity two-qubit Pauli operators.
Choosing $\lambda = \cfrac{16p}{15}$ gives us a combined probability of error equal to $p$ with each of the 15 two-qubit Pauli errors occurring with probability $\cfrac{\lambda}{16} = \cfrac{p}{15}$.
\item $\lambda_{\text{full}} = \cfrac{16}{15}$ is the maximum allowed parameter with the corresponding value of $p=1$ giving us an error channel where every two-qubit gate operation (such as C-NOT) is necessarily followed by a uniformly random two-qubit Pauli error.
\end{itemize}

In our experiments, we have used Qiskit’s Aer simulator to implement a custom noise model. First, each high-level quantum circuit was decomposed into elementary gates using Qiskit transpilation. Next, for each single-qubit and two-qubit gate, we injected depolarizing noise using Qiskit’s Quantum Error API, calibrated to the value of $p$. For example, setting $p = 0.02$ means each gate has a 2\% chance of being followed by a random Pauli error. We controlled the error budget by adjusting $p$ and observed the output probability distributions across 10,000 shots\footnote{Shots in quantum programming refers to the number of times the quantum algorithm is run on a quantum computer to get a probability distribution of the output states.} per run.

\subsection{Statistical Metrics}\label{sec:intuition}

The metrics that we use for analysis are defined on the outcomes of measurement operations
that are used to infer the solution of the computational problem. 
For ease of exposition, we work with measurement
operations in the computational basis of observables with non-degenerate eigenvalues. We will  refer to as eigenstates, the classical states in the computational basis to which the system collapses,  upon measurement. The technique that we discuss can easily be extended to other observables with degenerate eigenvalues.
\begin{enumerate}

\item \textbf{Most Probable States} ($MPS(r)$): 
As outcome of a measurement operation, the set of eigenstates that have probability masses within $r$\% of the probability of the highest probable eigenstate. In this work, we use $r=5$, and omit $r$ in the notation.

\item \textbf{Desired States ($DS$)}: The $MPS$ of a bug-free implementation of the quantum circuit of a quantum algorithm, run in a noise-free environment. Intuitively, these are the measurement outcome eigenstates that lead to the correct solution.

\item \textbf{Bias (\(\beta\))}: The total probability of measuring outcomes that do not belong to the set of desired eigenstates\footnote{Unless otherwise specified, by ``states" we will mean outcome eigenstates for the rest of the paper.}: 
\[
\beta=\sum_{i\notin DS} p_i,
\]
where $p_i$ is the probability of outcome being $i$.

\item \textbf{Entropy (\(S\))}: Used to quantify the uncertainty in measurement outcomes: 
\[
S = - \sum_{i=1}^{2^n} p_i \log_2(p_i),
\]
where $n$ is the number of qubits we are measuring.

\end{enumerate}

These metrics were selected because they collectively balance effectiveness and interpretability. Bias measures the deviation from desired outcomes, entropy captures the effect of noise on output uncertainty, and $MPS$ indicates whether the circuit’s dominant outcome matches expectations (i.e., the desired states). 
With these metrics, in the context of the output dilemma mentioned in section~\ref{sec:intro}, we can positively identify buggy implementation in the presence of 
noise below a threshold (please see section~\ref{sec:thresh}). When the implementation is correct, we can distinguish 
between the presence and absence of noise.

\section{Our Approach}\label{sec:Approach}
We begin this section by first discussing an example that inspires our approach---the ``Bias-Entropy Model''.
\subsection{A Motivating Example} \label{sec:motivating_example}
The effect of noise leads to uncertainty in the outputs of the quantum program, and running it over multiple shots will give us a probability mass function. To further study the possibility of a bug or noisy hardware, one could study this probability distribution to come to a conclusion.

\begin{figure}[h]
    \centering
    \begin{subfigure}{\linewidth}
        \centering
        \includegraphics[width=0.7\linewidth, height=0.45\columnwidth]{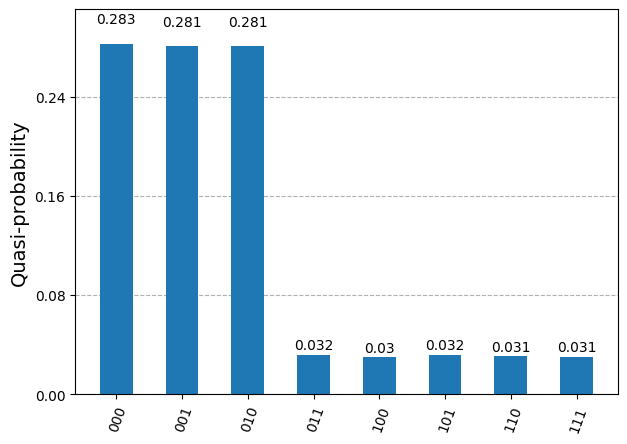}
        \caption{Grover: Bug-free and Noise-free}
        \label{fig:grover_bug_free_noise_free}
    \end{subfigure}
    \hfill
    \begin{subfigure}{\linewidth}
        \centering
        \includegraphics[width=0.7\linewidth, height=0.45\columnwidth]{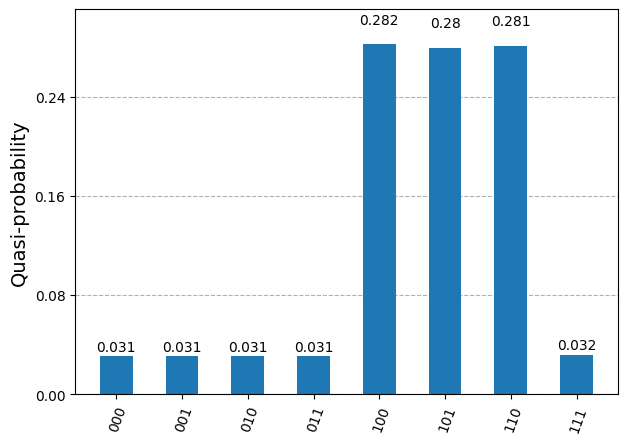}
        \caption{Grover: Buggy and Noise-free}
        \label{fig:grover_buggy_noise_free}
    \end{subfigure}

    \caption{Comparison of Grover's algorithm under different conditions: bug-free vs buggy.}
    \label{fig:grover_comparison}
\end{figure}

Bugs, as defined earlier, will lead to incorrect answers. This essentially means that buggy implementations lead to a scenario where $MPS$ does not match $DS$ as shown in Fig.~\ref{fig:grover_comparison}. Fig.~\ref{fig:grover_bug_free_noise_free} shows the quasiprobability
distribution plot generated by Qiskit for  a correct implementation of the Grover algorithm with $DS=\{000,001,010\}$. A bug will cause the $MPS$ to change to $MPS\neq DS$ as seen in Fig.~\ref{fig:grover_buggy_noise_free}.

\begin{figure*}
    \centering
    \begin{subfigure}{0.32\textwidth} 
        \centering
        \includegraphics[width=\textwidth, height=0.65\columnwidth]{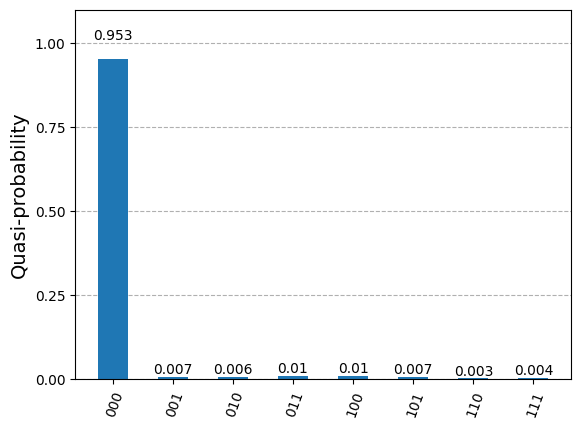}
        \caption{Noise-free}
        \label{fig:expected_outcome}
    \end{subfigure}
    \hfill
    \begin{subfigure}{0.32\textwidth} 
        \centering
        \includegraphics[width=\textwidth, height=0.65\columnwidth]{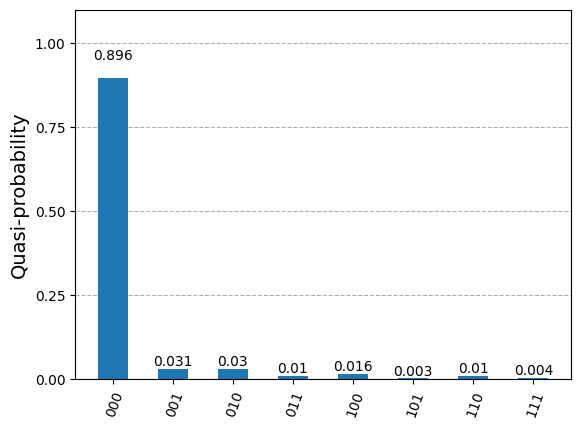}
        \caption{Slightly noisy}
        \label{fig:expected_outcome_little_noisy}
    \end{subfigure}
    \hfill
    \begin{subfigure}{0.32\textwidth}
        \centering
        \includegraphics[width=\textwidth, height=0.65\columnwidth]{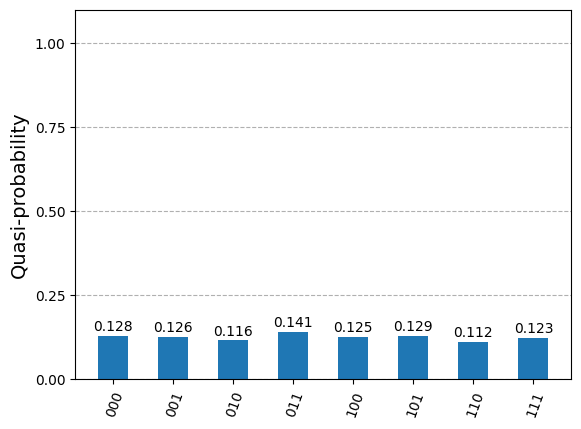}
        \caption{Highly noisy}
        \label{fig:expected_outcome_high_noise}
    \end{subfigure}

    \caption{Outcome of a correct implementation of Grover's algorithm under different noise levels}
    \label{fig:expected_outcome_all}
\end{figure*}

Fig.~\ref{fig:expected_outcome_all} shows the percentage of outcomes for a bug-free implementation of Grover's algorithm with DS = \{000\}.
One would expect the histogram given in Fig.~\ref{fig:expected_outcome}, where the expected correct state is the output for 95.3\% of the shots. However, the programmer may still be satisfied by the result in Fig.~\ref{fig:expected_outcome_little_noisy}, where the expected correct output is still the dominant state.
However, it would not be useful for the programmer to get a result like Fig.~\ref{fig:expected_outcome_high_noise}, where there is no discernible set of dominant states to reach a definitive answer.

All three figures mentioned above are run using the same piece of quantum code and on the same system. The only difference was the ``virtual environment" (simulator) they were run in. Fig.~\ref{fig:expected_outcome} was run in an ideal scenario where there is no noise. Fig.~\ref{fig:expected_outcome_little_noisy} was run in an environment with backend noise~\cite{qiskit_aer_noise_simulation}, a simplified noise model for a real device. Fig.~\ref{fig:expected_outcome_high_noise} was run in an environment using a custom noise model to simulate a highly noisy scenario.

The error introduced due to noise in each shot is different. Hence, it results in a spread of the probability distribution, taking probability mass away from the desired states and arbitrarily assigning this probability mass to other states (Section~\ref{sec:thresh}).

In summary, bugs will cause the $MPS$ of the probability distribution to change, whereas noise will only affect the spread of the probability mass function.
If the noise is not catastrophic, then the $MPS$ will remain unchanged.

\subsection{The Bias-Entropy Model}\label{sec:method}

In this section, we will introduce our approach to analyzing quantum noise and bugs. Throughout this paper we assume the noise remains constant across each shot. 
First, we give some useful definitions related to \emph{noise} that we will use.

\begin{bf} Definition 1 \end{bf} \emph{Gate Noise ($C$, $g$):} The probability of an erroneous output due to noise for gate $g$ of a quantum circuit $C$. 

\begin{bf} Definition 2 \end{bf} \emph{Noise Level ($C$):} the maximum value of \emph{Gate Noise($C$, $g$)} over all gates of the quantum circuit $C$.

For example, consider the following quantum circuit $C$:

\begin{center}
\begin{tikzpicture}[scale=0.7]
    \draw (0,0) -- (1,0);
    \draw (1,-0.5) rectangle (1.5,0.5);
    \node at (1.25,0) {$X$};
    \draw (1.5,0) -- (2.5,0);
    \draw (2.5,-0.5) rectangle (3,0.5);
    \node at (2.75,0) {$Y$};
    \draw (3,0) -- (4,0);
    \draw (4,-0.5) rectangle (4.5,0.5);
    \node at (4.25,0) {$Z$};
    \draw (4.5,0) -- (5.5,0);
\end{tikzpicture}
\end{center}

Further, let the noise be such that the \emph{Gate Noise ($C$, X)} is 3\%, \emph{Gate Noise ($C$, Y)} is 4\%, and  \emph{Gate Noise ($C$, Z)} is 5\%. Hence, in this scenario, the \emph{Noise Level ($C$)} is 5, which is the maximum of all individual gate noises. 

\begin{bf} Definition 3 \end{bf} \emph{Threshold Noise Level ($C$):} The least noise level at which $MPS \neq DS$ when a correct implementation of the quantum circuit $C$ is run on a noisy quantum computer.

This threshold is the catastrophic level of noise above which it will be impossible to distinguish
bugs from noise. \emph{Our results hold in the regime below such noise levels.} We will 
discuss in detail in Section~\ref{sec:thresh} how to estimate the thresholds.

As discussed in Section~\ref{sec:intro}, we have four cases of the output dilemma. Here, we will demonstrate how to quantify noise and bugs in those four cases, respectively.
\begin{enumerate}
    \item \textbf{No Bugs, No Noise}: In this case, we expect
    \begin{align*}
        \beta &\approx 0 \ \texttt{and} \\
        S &\approx\log_2(|DS|)
    \end{align*}
    Thus, the circuit will return $DS$ with the highest probabilities, thus $MPS = DS$.
    
    Moreover, the value of $\beta$ does not always need to be exactly zero, as quantum algorithms are inherently probabilistic. For example, in Grover's algorithm (Fig.~\ref{fig:expected_outcome}), the probabilities of all the marked states within $DS$ are relatively much higher than the probabilities of the states not in $DS$, which are close to zero. Since such an algorithm will result in almost equal probabilities of the desired states and that of the other states to be almost zero, the entropy would result in 
    \begin{align*}
        S&\approx-\sum_{i\in DS} \frac{1}{|DS|}\log_2(\frac{1}{|DS|}) \\
         &=-|DS|\frac{1}{|DS|}\log_2\big(\frac{1}{|DS|}\big)\\
         &=\log_2(|DS|).
    \end{align*}
    \item \textbf{Buggy, No Noise}: When there are bugs but negligible noise, the value of $\beta$ is expected to be high. When there is no external interference in a program, a bug will lead to measurement outcomes that are not in $DS$, i.e., $MPS \neq DS$.
    
    In such a scenario, we cannot comment on the entropy $S$ because when dealing with bugs, we could end up in any state. There could be bugs that give one wrong answer deterministically when the $|DS| > 1$, and there could also be cases where a bug may cause entirely random behavior, just like catastrophic noise. 
    \item \textbf{No Bugs, Noisy}: 
    Noise does not change $MPS$ if it is below \emph{Threshold Noise Level (C)} (see Section~\ref{subsec:noise_cause_bias}). Hence, $MPS$ would equal $DS$--all correct answers are in $MPS$. 
    
    \item \textbf{Buggy, Noisy}: Assuming noise levels below \emph{Threshold Noise Level ($C$)}: $MPS\neq DS$. A similar argument to the case of ``No Bugs, Noisy'' holds here as well, where a bug causes a change in $MPS$ leading to $MPS \neq DS$. The noise, being below \emph{Threshold Noise Level (C)}, will not change the $MPS$. This indicates the presence of a bug.
    
\end{enumerate}

This intuition yields Algorithm~\ref{alg:algorithm} that works in a regime of noise below \emph{Threshold Noise Level (C)}. We will discuss how to determine the threshold in the next subsection.

\begin{algorithm}
    \SetAlgoLined
    \KwIn{\text{\emph{Threshold Noise Level($C$)}}, measurements $\beta$, $S$, $MPS$, $DS$}
    \KwOut{Diagnostic result}
    
    \If{\text{\emph{Noise Level($C$)}} $< \text{\emph{Threshold Noise Level($C$)}}$}{
        \textbf{Subroutine:}\;
        Evaluate $\beta$, $S$, $MPS$, $DS$
        
        \uIf{$\beta \approx 0$ \textbf{and} $S \approx \log|DS|$}{
            \Return ``No bugs, No noise''\;
        }
        \ElseIf{$MPS = DS$}{
            \Return ``No bugs, Noise Present''\;
        }
        \Else{
            \Return ``Bugs present''\;
        }
    }
    \Else{
        \Return ``Noise too high''\;
    }
    \caption{Algorithm to differentiate between bugs and noise}
    \label{alg:algorithm}
\end{algorithm}

\subsection{Fixing Threshold Noise Level (C)}\label{sec:thresh}
Noise is a significant variable in current quantum computer hardware and leads to inconsistencies in the quantum measurement results \cite{cordier2025scalingproblemsizesenvironmental}. In simulations, we treat this noise as random errors to the circuit gates with arbitrary probability (Section~\ref{sec:Noise_model}). When this probability is very high, an error occurs after most of the gate operations in the circuit, which will lead to inconsistent answers when run across many shots (as shown in Fig.~\ref{fig:expected_outcome_high_noise}). We consider this condition of a system (hardware) as ``too noisy". It is not recommended to run the Quantum Program on the system under such conditions.

We now estimate a threshold of noise level, above which we classify the quantum computer to be too noisy and not suitable for experimentation at that point.

Let \(G\) denote the set of all gates other than measurement gates in the Quantum Circuit. In the example in the previous subsection, \(G = \{X, Y, Z\}\). \(|G|\) denotes the number of gates in the circuit.

To simplify the calculations for finding the threshold of noise level to label a system as too noisy, we need some assumptions:
\begin{enumerate}
    \item Whenever a gate is affected by noise, it will always lead to an anomalous outcome upon measurement, i.e., the effects of noise on multiple gates will not cancel each other out.
    \item Throughout the experiment, the noise level remains constant.
    \item In the absence of noise, each of the states in $DS$ occurs with a probability equal to $p=\cfrac{1}{|DS|}$.
\end{enumerate}

Let us define \(A_i\) as the indicator random variable as follows.

\[
A_i = \begin{cases}
1, & \text{if noise affects } G_i, \\
0, & \text{otherwise}.
\end{cases}
\]

We denote \emph{Noise Level ($C$)} by $P$.
Thus, the probability of an anomalous measurement outcome due to noise, that is, at least one gate being affected by noise, is: $ p( \bigcup_{i} A_i) \leq \sum_{i} p(A_i) \leq |G| P $, which follows from Boole's Inequality and the fact that $ \forall i$, $p(A_i) \leq P$.

To begin with, let the noise always lead to only one anomalous state. Hence, the bias due to noise becomes: $\cfrac{1}{|DS| + 1}$. That is, it causes the anomalous state to become a part of $MPS$. Moreover, the anomalous state and the states in $DS$ become equally likely with a probability of $\cfrac{1}{|DS|+1}$.

Therefore, $\tilde{P}^* = \frac{1}{(|DS|+1)|G|}$
would be a pessimistic estimate of \emph{Threshold Noise Level (C)}.

However, we find that in practice, an optimistic analysis is more 
appropriate. Therefore, we use the average case analysis below as \emph{Threshold Noise Level (C)}:
\subsubsection{Average Case Analysis}

In a general case, we have observed that in highly noisy systems, entropy is also high, which causes uncertainty in measurement, that is, all states are measured with almost equal probability.

As with our previous definition, \(|DS|\) denotes the number of states that are expected to be in the outcome of measuring the circuit, and we assume that each of the outcomes occurs with a probability of \(\cfrac{1}{|DS|}\) in the noise-free scenario.

However, for noise at or above \emph{Threshold Noise level(C)}, the probabilities of all states are observed to be almost equally likely, and this probability is $\approx \cfrac{1}{2^n}$, where
$n$ is the number of qubits.

Probability of an anomalous measurement outcome due to noise \(\leq |G|P^*\) (where $P^*$ denotes the noise level at the threshold for the average case).
But the Bias caused due to this noise is \(1 - \cfrac{|DS|}{2^n}\).

Equating the above two, we get the threshold of noise level for the average case:
\begin{displaymath}
P^* = \big(1-\frac{|DS|}{2^n}\big)\frac{1}{|G|} 
\stackrel{\text{def}}{\hbox{\equalsfill}}\textrm{\emph{Threshold Noise Level (C)}}.
\end{displaymath}

\section{Effects of Bias and Entropy on Random Quantum Circuits}\label{sec:effects_noise_bug}
In this section, we will show experimental observations on how bias and entropy are affected in the presence of bugs and noise separately. We will also study how the combined effect of bugs and noise would change the probability mass function of the measured states.

\subsection{Effect of noise on entropy} \label{subsec:effect of noise and entropy}
\textbf{Experimental setup:}
To identify the effect that noise would have on entropy, we have used noise models based on system snapshots (backend noise) of real quantum computers, provided by Qiskit~\cite{qiskit_fake_provider}. We have run several random circuits~\cite{qiskit_random_circuit} in simulators with backend noise and compared them with the results of noise-free simulation. 

According to research done by Ichikawa et al.~\cite{Ichikawa_2024}, the average number of qubits used for quantum computing is 10.3, and their median is 6.0. Thus, we use circuits of qubits in the range of 2 to 15. The authors further mention that the circuits used are shallow due to constraints posed by the current noise level. Therefore, for simplicity, we use depth in the range 1-5.

We chose to run our experiments on Qiskit fake backends (quantum machine simulators) because of the limited access and high costs of physical quantum computers. To simulate real quantum machine settings, we perform our experiments using 59 backend noise models run on a total of 3,560 random quantum circuits. 

\textbf{Observations:} In most of the cases, the entropy increases in the presence of a noisy backend. In the cases where entropy does not increase, all the possible states lie in $DS$, i.e. $|DS|=2^n$. Such a quantum program is of no significance and we will omit such cases in our study.

\begin{figure}[h]
  \centering
  \includegraphics[width=0.65\linewidth]{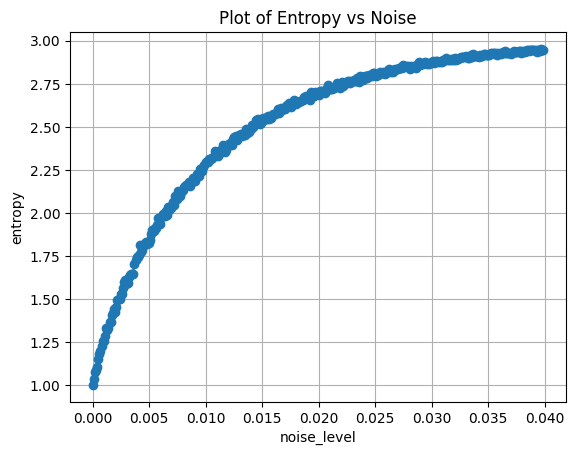}
  \caption{Effect of varying noise levels on a correct implementation of Grover's algorithm.}
  \label{fig:grover_noise}
\end{figure}

To explain the above observation, we use Fig.~\ref{fig:grover_noise}, which shows the effect of changing noise levels on entropy with a correct implementation of Grover's algorithm. The noise model that we employ is the custom noise model discussed in Section~\ref{sec:Noise_model}.
 As noise level increases, entropy also increases. As discussed in Section~\ref{sec:Noise_model}, this is due to the fact that the noise model randomly adds errors to the output state of each gate, which increases the weights of undesirable measurement outputs in the probability mass function. 
This increase will continue until the \emph{Threshold Noise Level ($C$)}. Beyond this, the graph saturates and reaches the maximum value of entropy for the given quantum circuit --- where we have mentioned all the states are empirically observed to have similar probability masses. From the graph, we observe the experimental threshold to be around 0.040. Comparing it with the theoretical threshold computed using $|DS| = 1$, $|G| = 51$ and $n=3$, we get $P^* = 0.015$. Although not a strong bound, one can check if the noise level is below this and before running the quantum program on the hardware.

\subsection{Effect of bugs on entropy}
\label{subsec:effect_bugs_entropy}
\textbf{Experimental Setup:} The effects of bugs on entropy were studied using randomly generated quantum circuits (random circuits). Buggy versions (mutants) of these circuits were then created using Muskit~\cite{Mendiluze2021}. A large number of mutants were created by adding extra gates into the circuit, and replacing and removing existing gates of the circuit. The entropies of these mutants were calculated and were then used to estimate a probability distribution. This was done in order to observe the distribution of entropy values when bugs are introduced into the circuit. This distribution of values is also compared with the entropy of the corresponding bug-free random circuit.

\begin{figure}[h]
  \centering
  \subfloat[Entropies concentrated away from the bug-free Entropy]{%
    \includegraphics[width=0.75\linewidth]{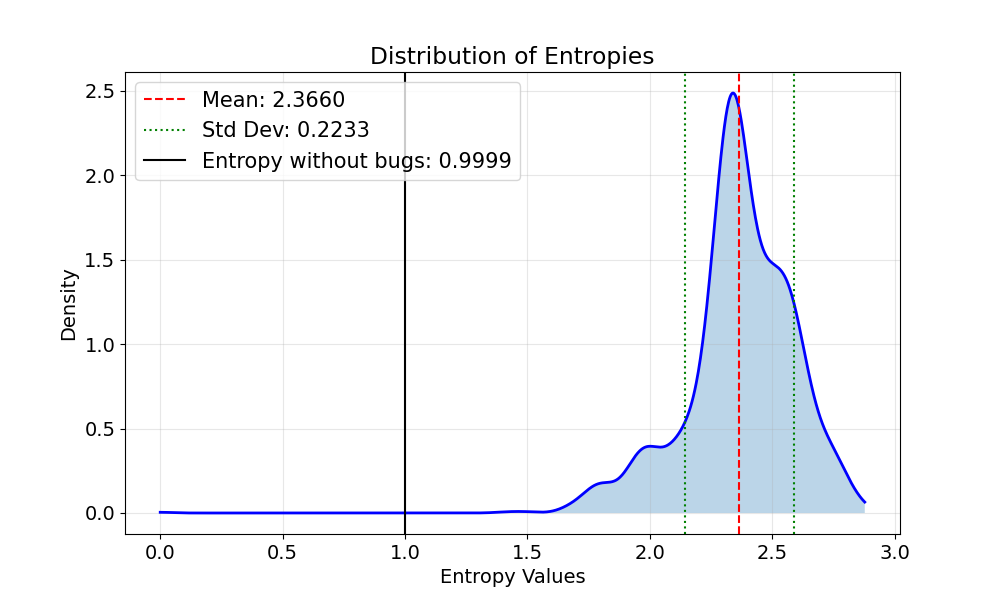}
  }\\[1ex]
  \subfloat[Entropies concentrated around the bug-free Entropy]{%
    \includegraphics[width=0.75\linewidth]{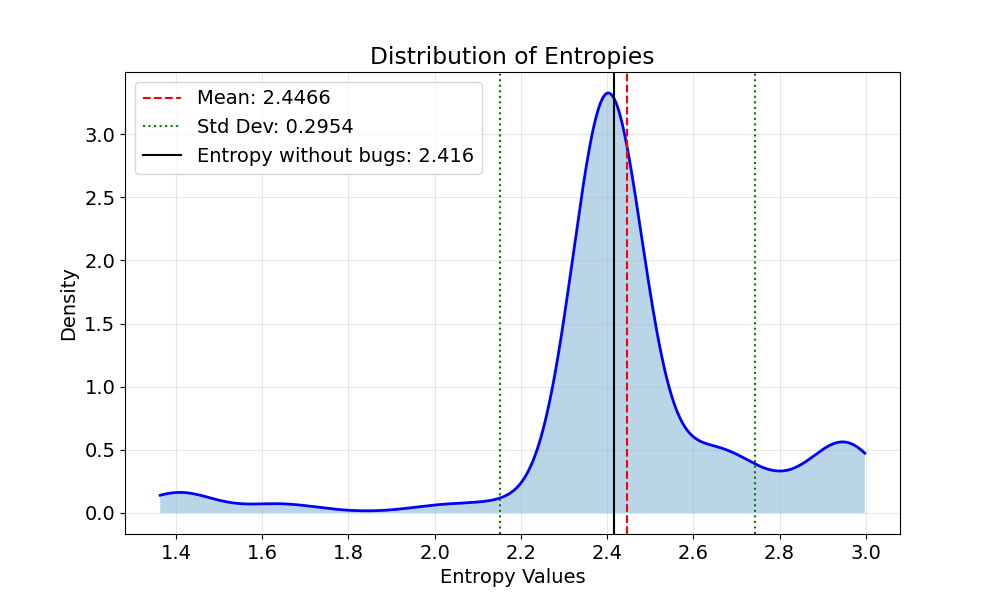}
  }
  \caption{Distribution of entropies with buggy circuits}
  \label{fig:Entropy_dist}
\end{figure}

\textbf{Observations:} The relationship between the entropy of a buggy quantum circuit and its corresponding bug-free version exhibits complex and unpredictable behavior, as illustrated in Fig.~\ref{fig:Entropy_dist}. This figure presents entropy distributions for two distinct circuits on 3 qubits, revealing markedly different patterns. The impact of bugs on entropy values varies significantly, both within a single circuit and across different circuits.
This is evident in the observation that the entropy deviation caused by a bug appears to be highly dependent on the specific characteristics of both the bug itself and the circuit in which it occurs. This variability makes it challenging to establish general rules or patterns regarding how bugs affect entropy values in quantum circuits.

Given the intricate nature of this relationship, we refrain from attempting to formulate any broad generalizations about the effects of bugs on entropy in quantum circuits. Instead, our findings suggest that each case requires individual analysis, taking into account the unique properties of the circuit and the nature of the introduced bug.

An additional noteworthy observation is the presence of a small but discernible density at the maximum possible entropy value for 3-qubit systems (i.e., 3). This indicates that a subset of bugs induces a state of complete randomness in the circuit's measurement output, resulting in an approximately uniform distribution across all 3-bit measurement states, similar to Fig.~\ref{fig:expected_outcome_high_noise}. However, the relatively low density at this maximum entropy point suggests that only a small fraction of bugs lead to this extreme effect.

\subsection{Effect of noise on bias} \label{subsec:noise_cause_bias}
\textbf{Experimental setup:}
The experimental setup remains the same as Section~\ref{subsec:effect of noise and entropy}:
We will use circuits with qubits in the
range of 2 to 15 having depth in the range of 1-5. We executed a series of random quantum circuits on simulators with backend noise and compared the outcomes with those from ideal, noise-free simulations.

\textbf{Observations:}
In almost every case except the one where all the states belong to $MPS$, we have observed that the random circuit with noise has more Bias than the same circuit without noise (ideal simulator). However, below the \emph{Threshold Noise Level ($C$)}, we observe that noise does not lead to a change in $MPS$.

\begin{figure}[h]
  \centering
  \includegraphics[width=0.65\linewidth, height=0.5\columnwidth]{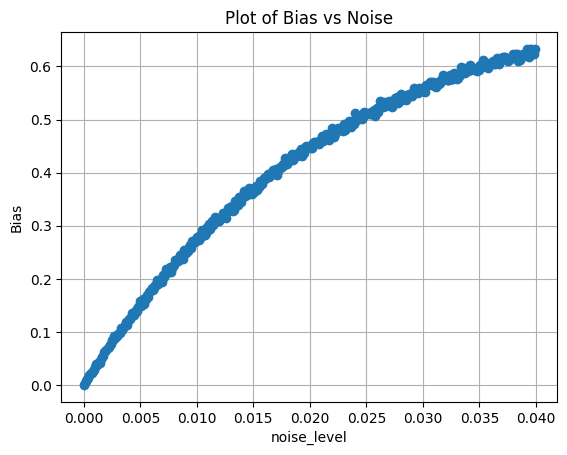}
  \caption{Noise Level vs Bias graph for Grover's algorithm}
  \label{fig:Bias_vs_Noise_Grover}
\end{figure}

To further see the effect of noise (again, as detailed in Section~\ref{sec:Noise_model}) on the bias, we can look at Fig. \ref{fig:Bias_vs_Noise_Grover}, 
which is a graph that plots the bias against varying noise levels for the same running example of Grover's algorithm. We can see from the figure that the bias gradually increases with an increase in noise until the \emph{Threshold Noise Level ($C$)}, where all the states become equally probable (Section~\ref{subsec:effect of noise and entropy}).

Tests on random circuits show similar behavior as Fig.\ref{fig:Bias_vs_Noise_Grover}, suggesting an increase in bias due to noise without affecting the $MPS$ is a general characteristic of quantum computations rather than being specific to any particular quantum algorithm or circuit design.

\subsection{Effect of bugs on bias}
\textbf{Experimental setup:}
The experimental setup remains the same as Section~\ref{subsec:effect_bugs_entropy}---buggy variants of random circuits generated using Muskit were compared against their corresponding unaltered circuits to measure the effect of bugs on bias in quantum programs.

\textbf{Observations:} 
In almost all of the mutants generated for a given random circuit, we see that the bias is either substantially higher or remains around the same. 
To see an example, we return to the Grover's algorithm.
In the non-buggy case for Grover's algorithm, we got bias to be equal to 0.0291 (almost 0 if not for the presence of noise). 

To investigate the effect of bugs, we introduce different combinations of Pauli and Hadamard gates on any of the three qubits
that we use for Grover search (unordered list size of $N=8$). 
In almost every buggy version of Grover's algorithm, the bias computed was found to be substantially higher than the bias of the correct implementation.
When these results are compared to those from our earlier experiment in Section \ref{subsec:noise_cause_bias}, it becomes evident that the increase in bias due to bugs is significantly greater than that caused by noise alone (below \emph{Threshold Noise Level ($C$)}). By definition, bias measures the total probability of obtaining outcomes that do not belong to the $DS$. Bugs tend to alter the $MPS$ of the circuit, leading to a condition of $MPS\neq DS$. This leads to a pronounced increase in bias, as the circuit is now more likely to produce incorrect outcomes. 
Moreover, when the noise is below \emph{Threshold Noise Level ($C$)}, $MPS$ will not be altered in the absence of a bug. Thus, this clearly explains why the increase in bias is not as substantial when it is caused only by noise.

\begin{table}[ht]
    \centering
    \resizebox{0.6\columnwidth}{!}{%
    \begin{tabular}{|c|c|c|}
        \hline
        Mutant No. & Bias & $\Delta$Bias \\
        \hline
        1 & 0.3951 & 0.3660 \\
        \hline
        2 & 0.9818 & 0.9527 \\
        \hline
        3 & 0.3865 & 0.3574  \\
        \hline
        4 & 0.9776 & 0.9484 \\
        \hline
    \end{tabular}
    }
    \caption{Comparison of bias for the bug-free code and buggy code of Grover's algorithm. The bias for the correct implementation is 0.0291.}
    \label{tab:Bug vs No bug bias}
\end{table}

Table~\ref{tab:Bug vs No bug bias} shows how bias changes in the presence of a bug for the various buggy versions of Grover's algorithm, given that the noise level is fixed. We can clearly see that the bias increases with the introduction of a bug. $\Delta$Bias in the table refers to the difference in bias for the mutant and the correct Grover implementation (0.0291).

\begin{figure*}[ht!]
    \centering
    \begin{subfigure}{0.32\textwidth}
        \centering
        \includegraphics[width=\textwidth, height=0.65\columnwidth]{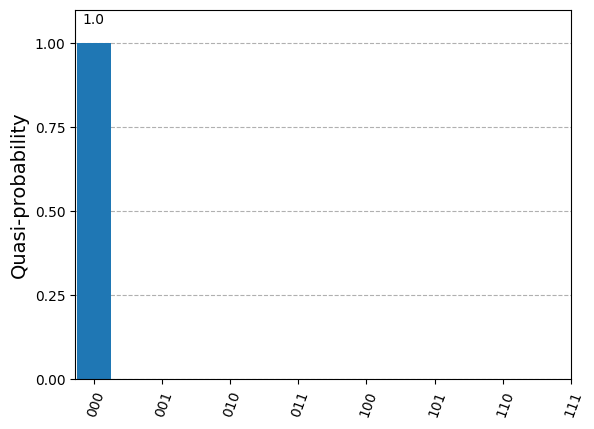}
        \caption{Noise-free, Entropy=0.0}
        \label{fig:DJA_cons_no_noise}
    \end{subfigure}
    \hfill
    \begin{subfigure}{0.32\textwidth}
        \centering
        \includegraphics[width=\textwidth, height=0.65\columnwidth]{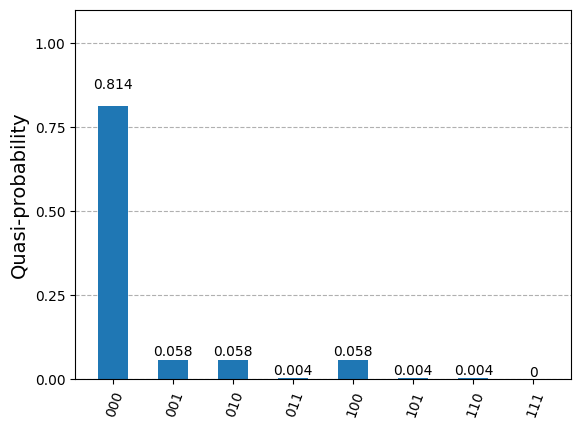}
        \caption{Noise below threshold, Entropy=1.0549}
        \label{fig:DJA_cons_below_thresh_noise}
    \end{subfigure}
    \hfill
    \begin{subfigure}{0.32\textwidth}
        \centering
        \includegraphics[width=\textwidth, height=0.65\columnwidth]{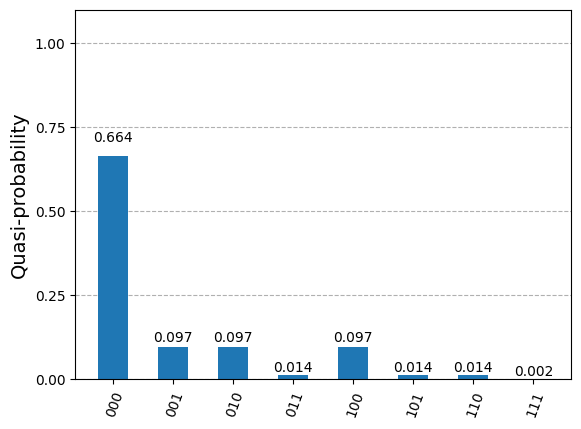}
        \caption{Threshold Noise, Entropy=1.535}
        \label{fig:DJA_cons_thresh_noise}
    \end{subfigure}
    \caption{Results for bug-free DJA with Constant functions}
    \label{fig:DJA_cons_results}
\end{figure*}
\begin{figure*}[ht!]
    \centering
    \begin{subfigure}{0.32\textwidth}
        \centering
        \includegraphics[width=\textwidth, height=0.65\columnwidth]{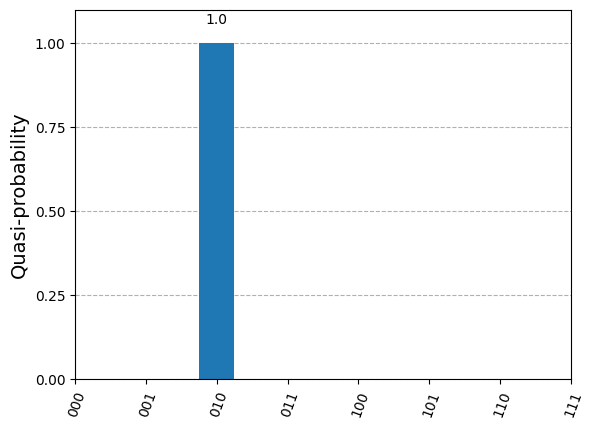}
        \caption{Noise-free, Entropy=0.0}
        \label{fig:DJA_cons_bug_no_noise}
    \end{subfigure}
    \hfill
    \begin{subfigure}{0.32\textwidth}
        \centering
        \includegraphics[width=\textwidth, height=0.65\columnwidth]{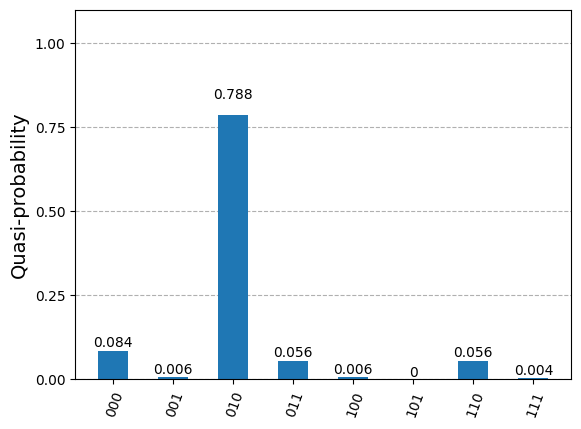}
        \caption{Noise below threshold, Entropy=1.1594}
        \label{fig:DJA_cons_bug_below_thresh_noise}
    \end{subfigure}
    \hfill
    \begin{subfigure}{0.32\textwidth}
        \centering
        \includegraphics[width=\textwidth, height=0.65\columnwidth]{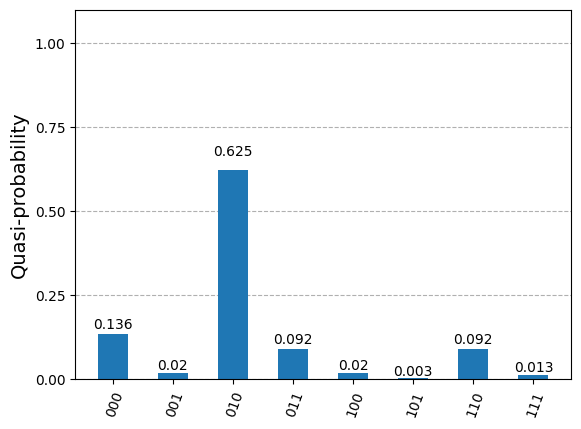}
        \caption{Threshold Noise, Entropy=1.7799}
        \label{fig:DJA_cons_thresh_noise}
        \label{fig:DJA_cons_bug_thresh_noise}
    \end{subfigure}
    \caption{Results for buggy DJA with Constant functions}
    \label{fig:DJA_cons_bug_results}
\end{figure*}

 \subsection{Analyzing the combined effect of bug and noise} \label{bug_plus_noise}
 Combining the ideas of the above subsections, a bug is detected by a deviation of $MPS$ from $DS$. On the contrary, noise does not alter the $MPS$ as long as \emph{Noise Level($C$)} is below \emph{Threshold Noise Level ($C$)}. Therefore, to decide if the algorithm is correct, one only needs to measure the $MPS$. Thus, applying Algorithm~\ref{alg:algorithm} on the probability mass function generated by performing multiple measurements on the quantum circuit of a program, one could distinguish whether the deviation in results is caused by a bug or by noise.

\section{Case Studies}\label{sec:case-study}
We have already discussed the Bias-Entropy approach as a running example in the context of the Grover search.
In this section, we investigate our approach for effectiveness and applicability on two other folklore algorithms: Deutsch-Jozsa algorithm (Section~\ref{subsec:DJA}) and Simon's algorithm (Section~\ref{subsec:simon}), respectively. 

We selected Grover’s, Deutsch–Jozsa, and Simon’s algorithms because they are among the most studied quantum algorithms with deterministic output profiles under ideal conditions. Their mathematical structures make them highly amenable to output distribution analysis (e.g., bias, entropy), allowing us to evaluate the robustness of our statistical analysis.
\begin{figure}[ht!]
    \centering
    \includegraphics[width=0.65\linewidth]{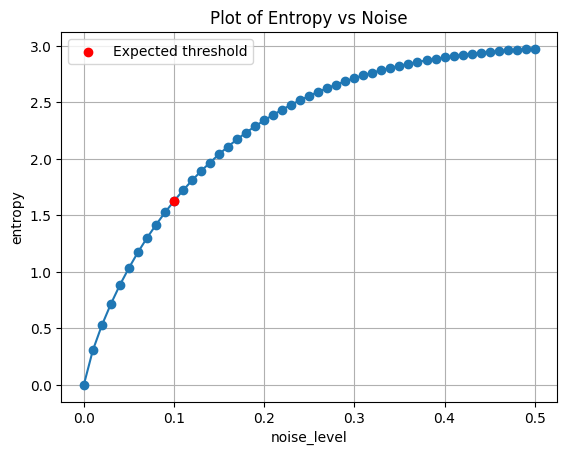}
    \caption{Entropy vs Noise-level for DJA-Constant}
    \label{fig:DJA_cons_EvN}
\end{figure}
\begin{figure*}[ht!]
    \centering
    \begin{subfigure}{0.32\textwidth}
        \centering
        \includegraphics[width=\textwidth, height=0.65\columnwidth]{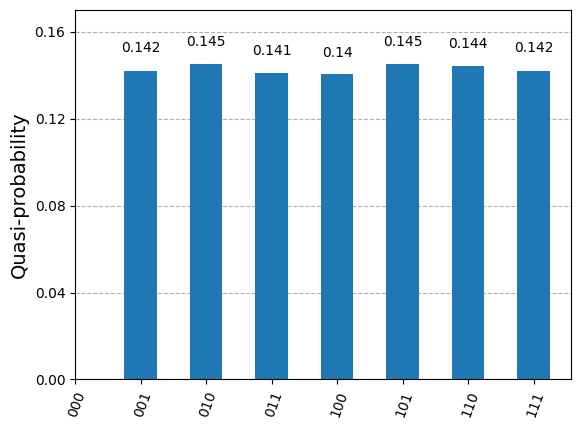}
        \caption{Noise-free, Entropy=2.8072}
        \label{fig:DJA_bal_no_noise}
    \end{subfigure}
    \hfill
    \begin{subfigure}{0.32\textwidth}
        \centering
        \includegraphics[width=\textwidth, height=0.65\columnwidth]{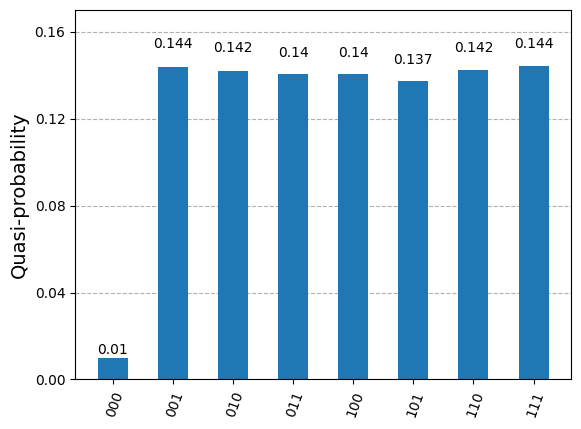}
        \caption{Noise below threshold, Entropy=2.8586}
        \label{fig:DJA_bal_below_thresh_noise}
    \end{subfigure}
    \hfill
    \begin{subfigure}{0.32\textwidth}
        \centering
        \includegraphics[width=\textwidth, height=0.65\columnwidth]{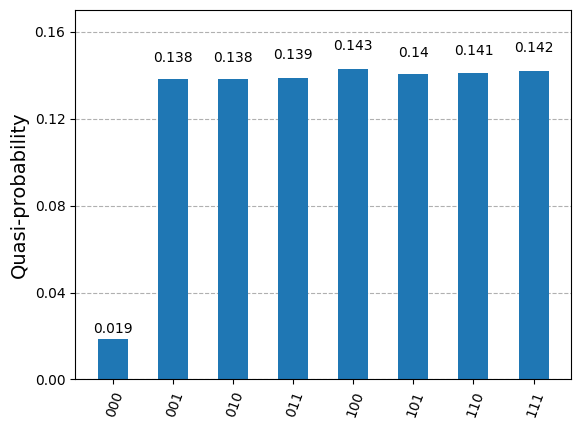}
        \caption{Threshold noise, Entropy=2.8887}
        \label{fig:DJA_bal_thresh_noise}
    \end{subfigure}
    \caption{Results for bug-free DJA with Balanced functions}
    \label{fig:DJA_bal_results}
\end{figure*}
\begin{figure*}[ht!]
    \centering
    \begin{subfigure}{0.32\textwidth}
        \centering
        \includegraphics[width=\textwidth, height=0.65\columnwidth]{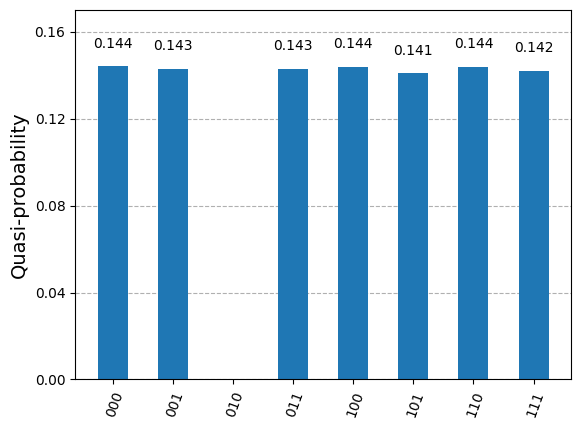}
        \caption{Noise-free, Entropy=2.8073}
        \label{fig:DJA_bal_bug_no_noise}
    \end{subfigure}
    \hfill
    \begin{subfigure}{0.32\textwidth}
        \centering
        \includegraphics[width=\textwidth, height=0.65\columnwidth]{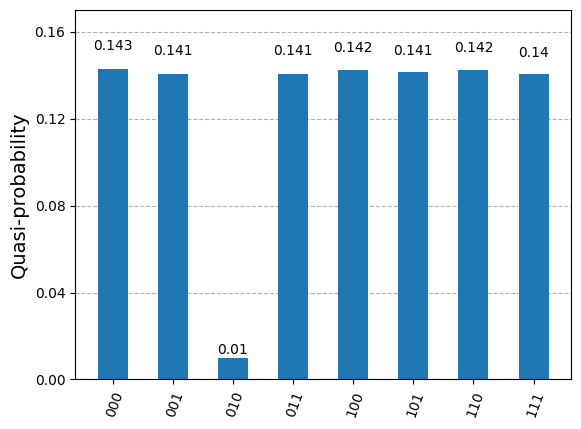}
        \caption{Noise below threshold, Entropy=2.8588}
        \label{fig:DJA_bal_bug_below_thresh_noise}
    \end{subfigure}
    \hfill
    \begin{subfigure}{0.32\textwidth}
        \centering
        \includegraphics[width=\textwidth, height=0.65\columnwidth]{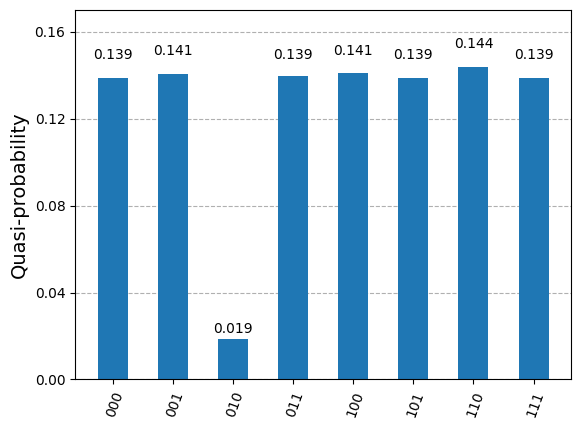}
        \caption{Threshold noise, Entropy=2.8889}
        \label{fig:DJA_bal_bug_thresh_noise}
    \end{subfigure}
    \caption{Results for buggy DJA with Balanced functions}
    \label{fig:DJA_bal_bug_results}
\end{figure*}
The desired states and run configurations for the implementations studied in this paper are as
follows. These configurations are chosen while keeping the run times and the consistency of results in mind.
\begin{enumerate}
    \item \textbf{Deutsch-Jozsa algorithm}: 
    The Deutsch-Jozsa algorithm is a quantum algorithm that solves the following problem. 
We are given a binary function oracle $f : \{0,1\}^n \rightarrow \{0,1\}$ with the promise that it is constant or balanced. We need
to decide whether the function is constant or balanced through queries
to the oracle. The Deutsch-Jozsa algorithm achieves this with a single query to the function, providing an exponential 
speedup over classical algorithms.
    
    For the Deutsch-Jozsa algorithm, $DS=\{0^n\}$ for a constant function, and $DS=\{0,1\}^n - \{0^n\}$ for a balanced function, where $n$ is the length of the input states of the function to be classified as constant or balanced. 

    \item \textbf{Simon's algorithm}: 
    Simon's algorithm is a quantum algorithm that solves the following problem in polynomial time. We are given a function oracle $f:\{0,1\}^n\rightarrow\{0,1\}^m$ for $m\leq n$
    such that $f(x)=f(x')$ if and only if $x'=x\oplus s$ for a hidden ``bit mask" $s$. We are required to find $s$ through queries to the $f$ oracle. 

    In Simon's algorithm, $DS=\{ y \in \{0, 1\}^n \mid (y \cdot s) \bmod 2 = 0 \}$, where $s$ is the bitmask string and $n=|s|$. Therefore, $N=2^n$ is the size of the domain of the promise function. 

\end{enumerate}

\subsection{Deutsch-Jozsa Algorithm}\label{subsec:DJA}
We now do the experiments for a small instance of Deutsch-Jozsa.
We run 10,000 shots on the simulator with 10,000 randomly generated promise functions with $n=3$.
For this experiment, we introduce a bug in the form of an extra Pauli-X gate on the second qubit.

\subsubsection{For Constant Functions}\label{subsubsec:DJA-cons}
Recall that a function is said to be constant if it outputs the same answer (either 0 or 1) for any binary string.
The parameters for the Deutsch-Jozsa Algorithm (DJA) (for constant functions) to calculate the theoretical bound are as follows:
\begin{itemize}
    \item \(n\) = 3
    \item \(|DS|\) = 1
    \item $|G|$ = 8.5 (averaged over 10000 constant functions)
\end{itemize}

Applying the formula for the theoretical bound, we get $P^* = 0.103$.

Checking the value of entropy for this bound in Fig.~\ref{fig:DJA_cons_EvN}, we see that it is significantly less than the point at which the curve flattens.

The histograms for bug-free implementations of DJA for constant functions are shown in Fig.~\ref{fig:DJA_cons_results}. As we can see, $MPS$ remains equal to $\{000\}$, which is equal to $DS$.

For the buggy implementation of DJA for constant functions, Fig.~\ref{fig:DJA_cons_bug_results} shows that for all three cases of noise less than or equal to the threshold, $MPS\neq DS$, since $DS=\{000\}$. 
\begin{figure}[ht!]
    \centering
    \includegraphics[width=0.65\linewidth]{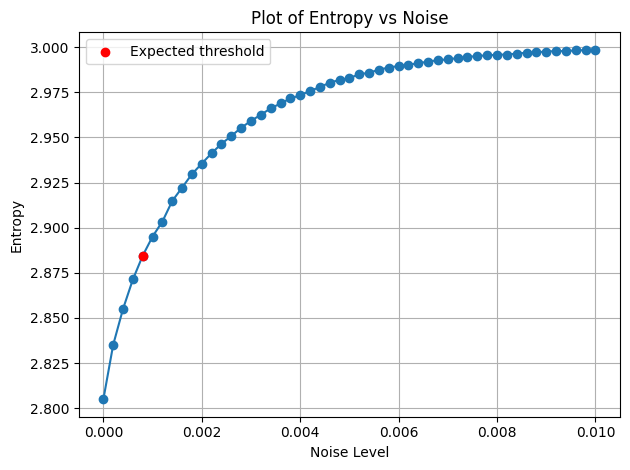}
    \caption{Entropy vs Noise-level for DJA-Balanced}
    \label{fig:DJA_bal_EvN}
\end{figure}
\begin{figure*}[ht!]
    \centering
    \begin{subfigure}{0.32\textwidth} 
        \centering
        \includegraphics[width=\textwidth, height=0.65\columnwidth]{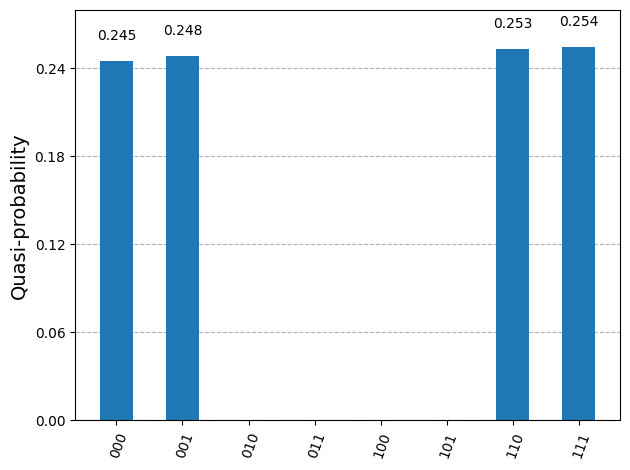}
        \caption{No Noise, Entropy=1.999}
        \label{fig:simon_noiseless_histogram}
    \end{subfigure}
    \hfill
    \begin{subfigure}{0.32\textwidth} 
        \centering
        \includegraphics[width=\textwidth, height=0.65\columnwidth]{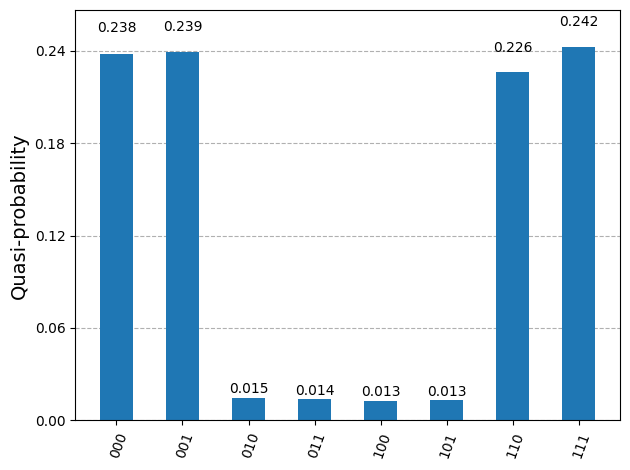}
        \caption{Noise below threshold, Entropy=2.305}
        \label{fig:Simons_Noise_0.003}
    \end{subfigure}
    \hfill
    \begin{subfigure}{0.32\textwidth} 
        \centering
        \includegraphics[width=\textwidth, height=0.65\columnwidth]{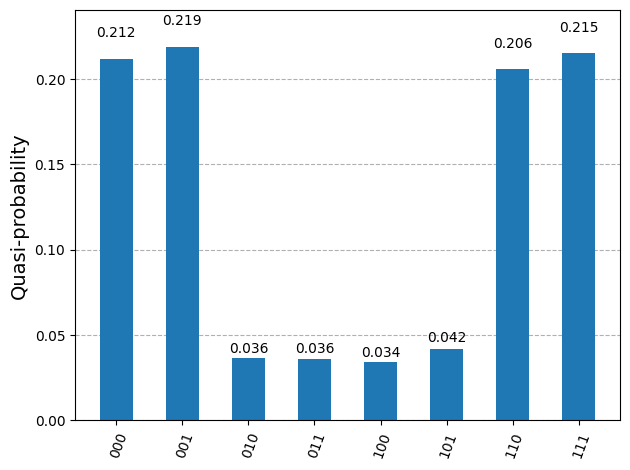}
        \caption{Threshold noise, Entropy=2.605}
        \label{fig:simon_threshold_histogram}
    \end{subfigure}
    \caption{Results for bug-free implementation of Simon's Algorithm at various noise levels}
    \label{fig:Simon_Results}
\end{figure*}
\begin{figure*}[ht!]
    \centering
    \begin{subfigure}{0.32\textwidth}
        \centering
        \includegraphics[width=\textwidth, height=0.65\columnwidth]{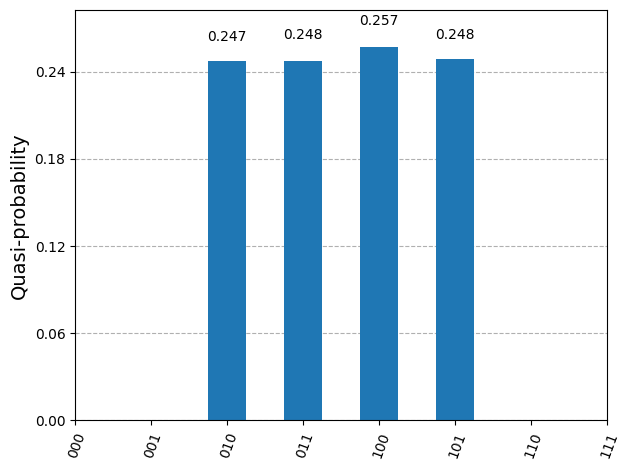}
        \label{fig:Simon_bug_noiseless}
        \caption{No noise, Entropy=1.999}
    \end{subfigure}
    \hfill
    \begin{subfigure}{0.32\textwidth}
        \centering
        \includegraphics[width=\textwidth, height=0.65\columnwidth]{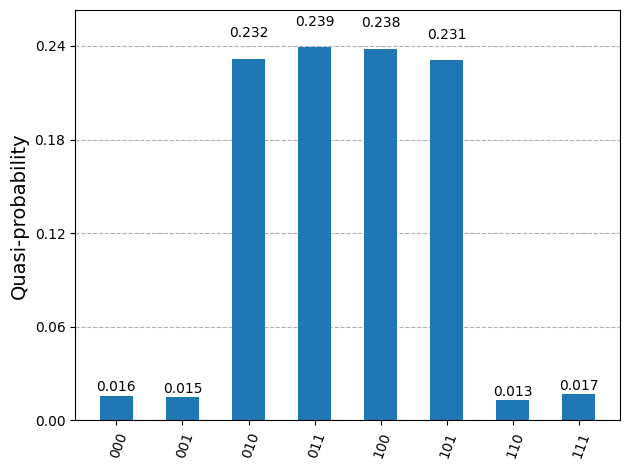}
        \label{fig:Simon_bug_lowNoise}
        \caption{Noise below threshold, Entropy=2.359}
    \end{subfigure}
    \hfill
    \begin{subfigure}{0.32\textwidth}
        \centering
        \includegraphics[width=\textwidth, height=0.65\columnwidth]{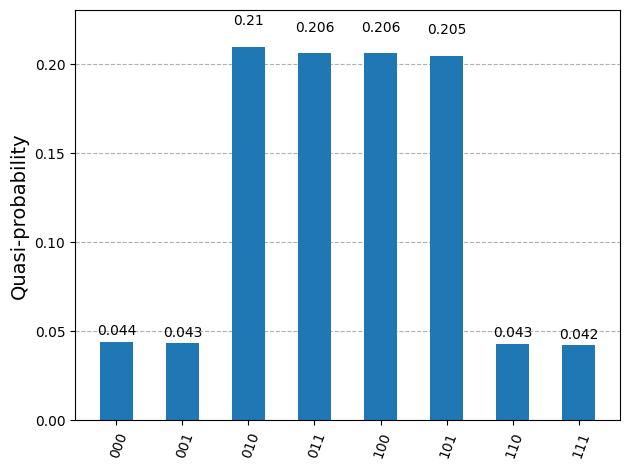}
        \label{fig:Simon_Bug_thresh_noise}
        \caption{Threshold Noise, Entropy=2.606}
    \end{subfigure}
    \caption{Results for a buggy implementation of Simon's Algorithm}
    \label{fig:Simon_buggy_results}
\end{figure*}

\subsubsection{For Balanced Functions}\label{subsubsec:DJA-bal}
A function is said to be balanced if it outputs 0 for exactly half of all binary strings in its domain and 1 for the other half.

The parameters for DJA (for balanced functions) to calculate the theoretical bound are as follows:
\begin{itemize}
    \item \(n\) = 3
    \item \(|DS|\) = 7
    \item \(|G|\) = 144.5 (averaged over 10000 balanced functions)
\end{itemize}

Using these, we get $
P^* = 0.00087$.
Checking the entropy value for this bound in Fig.~\ref{fig:DJA_bal_EvN}, we see that it is significantly less than the point of maximum entropy.

The histograms for bug-free implementations of DJA for balanced functions are shown in Fig.~\ref{fig:DJA_bal_results}. As we can see, $MPS$ remains equal to the set of all possible states except $000$.

For the buggy implementation of DJA for balanced functions, Fig.~\ref{fig:DJA_bal_bug_results} shows that for all three cases of noise less than or equal to the threshold, $MPS\neq DS$, since $000 \in MPS$.

\subsection{Simon's Algorithm}\label{subsec:simon}

\begin{figure}[ht!]
    \centering
    \includegraphics[width=0.60\linewidth, height=0.5\columnwidth]{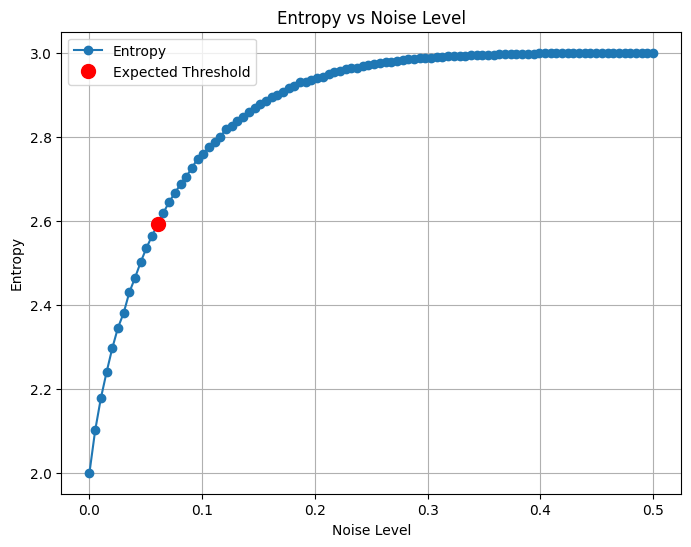}
    \caption{Entropy vs Noise curve for Simon's algorithm}
    \label{fig:Simons_Entropy_vs_Noise.png}
\end{figure}
We run  10,000 shots on the simulator with $n=3$.
For our experiment, we have used `110' as the string s. The following are the parameters used to calculate the theoretical threshold. 

\begin{itemize}
    \item \(n\) = 3
    \item \(|DS|\) = 4
    \item \(|G|\) = 8
\end{itemize}
Using these, we get
$P^* = 0.0625$
which is below the experimentally observed bound (0.30), which is the point at which the Entropy vs Noise curve flattens as shown in Fig. \ref{fig:Simons_Entropy_vs_Noise.png}. Fig.~\ref{fig:Simon_Results} shows the results of running Simon's algorithm on the backend at various different noise levels. From Fig. \ref{fig:simon_noiseless_histogram}, we measure $\beta=0$ and the entropy, $S=1.999\approx log_2|DS|$. Thus, the Algorithm \ref{alg:algorithm} would rightly return ``No bugs, No noise''. Fig. \ref{fig:Simons_Noise_0.003} shows the result when Simon's algorithm is run on a noise model with noise level less than the threshold. We get $S=2.305$, which does not match the first conditional in Algorithm \ref{alg:algorithm}. However, since we can clearly see that $MPS=DS$, the algorithm would return ``No bugs, Noise present''. 
Fig. \ref{fig:Simon_buggy_results} shows the distribution of a buggy implementation of Simon's algorithm. Specifically, the bug that we introduced was application of Pauli-X gates on the second qubit before performing the final measurement. In this case, the bias  was $\beta=0.931$ and entropy $S=2.359$. Moreover, we can see that the $MPS$ has changed and does not match the $DS$, that is, $MPS\neq DS$. Thus, the algorithm returns ``Bugs Present".

\section{Threats to Validity}\label{sec:threats}
We now summarize some potential threats to validity of the work reported in this paper.

\textbf{Internal Validity.} Our experiments are based on depolarizing noise models available in simulators (Section~\ref{sec:Noise_model}). While widely adopted, this model may not capture all characteristics of physical devices, potentially affecting the generalizability of our threshold estimates. To mitigate this threat, we incorporate both custom depolarizing models (Section~\ref{sec:Noise_model}) and Qiskit’s backend noise models (Section~\ref{subsec:effect of noise and entropy}), which emulate real-device noise profiles (e.g., IBMQ qubit decoherence). This hybrid approach captures a broader range of stochastic errors than pure depolarizing channels.

Moreover, we use Muskit to inject quantum bugs, which are mutated gates but may not cover all bug types (e.g., algorithmic design flaws or timing errors). We diversify mutations across 3,560 random circuits (Section~\ref{sec:effects_noise_bug}), covering common gate-level errors. While algorithmic bugs remain underrepresented, our empirical analysis confirms bug-induced bias/entropy shifts across diverse mutants.

\textbf{External Validity.} Our validation focuses on Grover, Deutsch-Jozsa, and Simon’s algorithms. While these are foundational, our findings may not generalize to probabilistic algorithms (e.g., variational quantum eigensolver or quantum approximate optimization algorithm).
As an initial study, we select foundational algorithms with well-defined $DS$ and noise responses, establishing a baseline for statistical metrics. We will extend our methods to more non-deterministic quantum algorithms in future work.

In addition, our experiments mostly use shallow circuits (depth 1–5, $\leq$15 qubits). Larger circuits with deeper entanglement may exhibit compounded noise/bug interactions beyond our threshold model. However, the current circuit dimensions are aligned with NISQ-era averages~\cite{Ichikawa_2024}. Our experiments on 3,560 random circuits further validate trends across shallow architectures typical of current quantum software. We plan to explore more validations as more resources become available.

\section{Conclusions and Future Studies}\label{sec:conclusions}
We introduced a robust statistical methodology designed to distinguish between quantum software bugs and hardware noise.
 Leveraging Bias, Entropy, and Most Probable States as probabilistic metrics, our approach provides quantum software developers with clear diagnostic insights into unexpected quantum program behaviors. Empirical studies using folklore quantum algorithms validated our methodology, highlighting its efficacy and applicability in practical quantum programming scenarios.
 
Future work will aim to extend this statistical framework to more complex quantum algorithms and larger quantum circuits. 
Experiments on actual quantum computers will offer further insight into the effectiveness of the technique.
Specifically, the quantum computer will allow working with and studying the effect of noise on bias and entropy.
Finally, we expect that trying out a wide variety of bugs would help in fine-tuning the bias-entropy characterization
towards identifying the bugs, if present.
We also plan to explore automated diagnostic tools integrating machine learning techniques to further enhance accuracy and scalability. Incorporating adaptive noise modeling and real-time diagnostic capabilities could significantly advance debugging efficiency, addressing emerging challenges as quantum technology scales.

\bibliographystyle{IEEEtran}
\bibliography{arxiv}

\end{document}